\documentstyle[11pt,newpasp,twoside,psfig]{article}
\def\edcomment#1{\iffalse\marginpar{\raggedright\sl#1\/}\else\relax\fi}
\reversemarginpar
\begin{document}
\title{$Chandra$ Observations of Embedded Young Stellar Objects}
\author{Katsuji Koyama}
\affil{Department of Physics, Graduate School of Science, Kyoto University,
Kita-Shirakawa, Sakyo-ku, Kyoto, 606-8502, Japan}
\begin{abstract}
This paper reviews the results of the $Chandra$ deep exposure
observations on star forming regions.
 The $\rho$ Ophiuchi ($\rho$ Oph) cloud cores B-F 
reveal $\sim$ 100 X-ray sources above the detection limit of 
$\sim$10$^{28}$ ergs s$^{-1}$.  About 2/3 of the X-ray sources are 
identified with an optical and/or infrared (IR) object, including 
significant numbers of class I sources.  The class I sources exhibit
higher temperature plasma than those of class II and III sources. These features
are confirmed with a larger number of X-ray sources in the Orion molecular clouds
2 and  3  (OMC 2 and 3). 

Hard X-ray emissions  are found from the sub$mm-mm$ dust cores, MMS 2 and MMS 3 
in the northern part of  OMC 3. 
These cores show  outflows  in the radio and IR bands, hence are in a
very early 
phase of star formation, possibly in the class 0 phase.  The spectra are heavily
absorbed suggesting that the X-ray sources are embedded in the cores. 

The giant molecular cloud Sagittarius B2 (Sgr B2) exhibits more than dozen X-ray
sources, two of which are  associated with the HII complex Sgr B2 Main (the ultra compact 
HII regions F-I). These show an absorption of $\gg$10$^{23}$ Hcm$^{-2}$, which is the largest among the 
known stellar X-ray sources.
 The  X-ray source in the HII regions F-I shows strong  
K-shell transition lines from  He-like and neutral irons, while that in the east of the 
HII region I has only a weak  line. 
No strong X-ray emission is  found from any  other HII complexes; hints of
weak X-rays are found from only  Sgr B2 North (the HII region  K) and South (H).

 The Monoceros R2 cloud exhibits X-ray emissions from young high-mass stars.
IRS 1,  possibly in  zero-aged main-sequence,  shows rapid time variability   
and a thin thermal spectrum  of $\sim 2$ keV temperature. 
Similar X-ray behaviors are found from younger 
high-mass stars, IRS 2 and IRS 3. These X-ray features are in contrast to
the low temperature plasma ($\leq$ 1 keV) and moderate variability 
found  in  high-mass main-sequence stars.

X-rays from 7 brown dwarfs and candidates are found 
in the  $\rho$ Oph cloud cores A--F, which comprise 40\% of those selected
with the IR observations. The X-rays are variable with occasional flares.
The X-ray spectra are fitted with a thin thermal plasma  of 1-3 keV,
with the luminosity ratio of X-ray to bolometric of $10^{-3}-10^{-5}$.  
These properties are essentially the same as those of low-mass pre-main sequence 
and dMe stars.
\end{abstract}
\section{Introduction}
Young stellar objects (YSOs) evolve from  molecular cloud
cores through protostars and pre main-sequence (PMSSs)  to zero-aged  main-sequence stars (ZAMSSs).
The  evolutional track  of low mass YSOs  can be well traced with the 
infrared (IR)-{\it mm} band spectra. 
Class 0 stars (class 0s), the early phase of protostars, show the spectral  peak at the {\it mm} 
band,  arising from the gas envelope. 
Class I stars (class Is), the later phase of protostars,  exhibit the emission peak
at the  mid- to far-IR wavelengths, and the spectra are still dominated by the accreting gas envelope.  
T Tauri stars (TTSs) are the quasi-static mass-condensation phase after the
dynamical mass accretion phase of protostars,  
and are divided into class IIs and IIIs; the formers
correspond to classical T Tauri stars (CTTSs), which  have a gas disk and show the near- to mid-IR excess 
emission from the disk. When the accretion disk disappears, CTTSs evolve to weak-lined T Tauri stars (WTTSs) 
or class IIIs, which have a black body
spectrum from the stellar photosphere with the peak at near-IR.
This phase lasts until hydrogen burning is ignited, which is the phase of ZAMSSs.  

TTSs (class IIs and IIIs) have been  known to exhibit fairly strong X-rays with occasional flares.  
The X-ray spectra are described with a thin thermal  plasma model of 
a temperature ranging from  1 to a few keV.  The X-rays are similar to, 
but have a larger flux and a higher temperature than those of the sun,
which leads to the general consensus that 
the X-ray origin is  due to the enhanced  solar-type magnetic activity.
The X-rays become less active as the 
low mass stars evolve to main sequence stars (MSSs) (for reviews, see  Feigelson \& Montmerle 1999). 

X-rays from class I protostars are discovered with either the hard X-ray imaging 
instrument on board $ASCA$ (Koyama et al. 1996, Kamata et al. 1997) 
or the deep exposure observations with $ROSAT$ 
(e.g. Grosso et al. 2000).  The X-ray features are similar  to those of more evolved 
stars, like TTSs, hence would be also due to the solar-type magnetic activity.
However, more quantitative study on protostellar X-rays in comparison with TTSs
has not be done, due to very limited  samples of protostars; the
X-ray detection rate from class Is was only 10\%, 
no X-ray sample from class 0 protostars, the youngest  phase of the stellar evolution, 
has been found. 

Since the evolution of high mass YSOs is very  rapid,
the evolutional classifications (or phase) are not clear-cut. 
Also the X-ray study on high mass YSOs is fur behind those of low mass stars, because of limited  samples.
The $ASCA$ satellite has found  hard X-rays from the center of 
giant molecular clouds (GMCs), the site of 
high-mass star formation (e.g. Yamauchi et al. 1997; Hofner \& Churchwell 1997; 
Sekimoto et al. 2000; Hamaguchi, Tsuboi, \& Koyama 2000).  However, the limited 
spatial resolution of $\sim 1'$, did not allow us to uniquely  resolve 
high mass YSOs in the GMCs, due to their relatively large distance and 
high density of stellar populations.  

Stellar X-rays have appeared well before the Hydrogen burning phase (MSSs), 
regardless of the stellar mass.  Therefore it may be conceivable that brown dwarfs, 
the sub-stellar object with the mass below the hydrogen burning limit of 
0. 08$M_\odot$ , also emit X-rays. In fact, $ROSAT$ discovered  X-ray emission 
from brown dwarfs (Neuh\"auser \& Comer\'on 1998).  However the number of,  and  
the spectrum/timing information of, the X-ray detected blown dwarfs  are still poor.  

As mentioned above, most of the previous  X-ray studies on YSOs have been concentrated 
on the low mass  stars in a TTS phase (class IIs and IIIs). 
This paper extends the   X-ray frontier toward three directions, using
the hard X-ray instruments with a fine spatial resolution of $Chandra$:
the X-ray emission from high mass YSOs,
protostars (class 0s and Is), and young blown dwarfs (BDs).  
The study is based on the recent $Chandra$  results.  
The star forming regions referred in this paper  are:  $\rho$ Ophiuchi ($\rho$ Oph)
, Orion Molecular Clouds 2 and 3 (OMC 2 and OMC 3), Sagittarius B2 (Sgr B2) 
and Monoceros R2 (Mon R2).
Detailed and individual reports  are: Imanishi, Koyama \& Tsuboi (2001); 
Tsuboi et al. (2001); Tsujimoto et al. (2001a, b); 
Takagi, Murakami \& Koyama (2001); Kohno, Koyama \& Hamaguchi (2001); 
and Imanishi, Tsujimoto \& Koyama (2001).

\section{Low Mass Protostars}
In the $Chandra$  ACIS-I observation on the $\rho$ Oph
cores B-F, we detect $\sim 100$ X-ray sources above the threshold of 
$\sim10^{28}$ ergs s$^{-1}$, of which  18
are identified with class Is \footnote {Class I
sources are defined by the near- to mid-IR band or near- to far-IR band
spectra (Andr\'e \& Montmerle 1994; Casanova et al. 1995; Chen et
al. 1995; Chen et al. 1997; Motte, Andr\'e, \& Neri 1998; Luhman \&
Rieke 1999; Grosso et al. 2000). Since the classifications in these
papers are not fully consistent with each other, I simply regard  a class I 
if any of the papers referred to be a class I source}.  
In the same field, we find 26 class Is in the IR-band star catalogs.
Thus about 70\% of the IR-selected class Is are X-ray sources,
which is higher ratio than that  with $ROSAT$ PSPS of about 
$\sim$10\% (Carkner, Kozak, \& Feigelson 1998).
The high detection rate of class Is ($\sim$70 \%)  with  {\it
Chandra} is  primary  due to the high sensitivity, especially in the hard X-ray 
band.  

Having reasonable numbers of X-ray samples of low mass YSOs
(class I - III), we fit the X-ray spectra with a thin thermal plasma model
for all the bright X-ray sources, 
then  find that the X-ray temperatures 
and absorptions of class Is  are generally larger than those 
of class IIs and IIIs.  
Also X-ray flares from  class Is show a hint of higher duty ratio,  
temperature  and  luminosity, than  those from more evolved class II and III
stars. 

A remarkable finding is a giant flare from YLW 16A with the peak luminosity
of $\sim 10^{31}$~ergs~s$^{-1}$. At the flare, the temperature rises
to $\sim10$ keV with the 6.7 keV line emission from He-like irons. We also find
the 6.4 keV line of  equivalent width $\sim100$ eV, which would be fluorescence from cold irons 
in the circumstellar 
gas. Since no time-lag is found between the 6.4 keV line and continuum flux, the
site of fluorescence should be very near the protostar, possibly the accretion disk.  
The large equivalent width favors  a disk face-on geometry.
Likewise, the 6.4 keV line is  a powerful diagnostics for the structure of the gas
envelope of YSOs.

A larger number of X-ray emitting low-mass YSOs is obtained from  OMC 2 and 3.
$Chandra$ detects $\sim 400$ X-ray sources in a $17'\times17'$ field
of the clouds. Our unified spectral analyses for all the X-ray bright
sources ($\sim 120$) reveal similar X-ray features as those   
found in the  $\rho$ Oph molecular cloud;
X-ray temperatures and absorptions  of class Is  are generally 
larger than those of class IIs and IIIs (Tsujimoto et al. 2001a).  
\begin{table}
\begin{center}
\begin{tabular}{cccc}
\multicolumn {4} {c} {Table 1. X-rays from Class 0 Candidates}\\ \\
\hline \hline
Position$^1$ &$N_{\rm H}$ $^2$  &$L_{\rm x}$ $^3$  &Associated sources$^4$  \\
\hline              
$18''.22, -34''.0$	&25(12-54) 	&2      & MMS 2, CSO 6, H$_2$ flow B,\\
						&&&HH 331, VLA 1, H$_2$ Jet\\
\multicolumn {4} {r} {- - - - - - - - - - - - - - - - -} \\ 	
$18''.93, -51''.0$	&14(7-34) 	&1	& MMS 3, CSO 7, H$_2$ flow D	\\
\hline  \\
\multicolumn {4} {l} {Temperature and abundance are fixed to be 3.2 keV and 0.3 solar,}\\
\multicolumn {4} {l} {respectively.  Parentheses indicate  90\% confidence range.}\\
\multicolumn {4} {l} {(1) Off set = $(05^h 35^m,-05^{\circ} 00')_{J2000}$} \\
\multicolumn {4} {l} {(2) Absorption in units of 10$^{22}$ Hcm$^{-2}$}\\
\multicolumn {4} {l} {(3) The 0.5-10 keV band luminosity in units of $10^{30}$~ergs~s$^{-1}$}\\
\multicolumn {4} {l} {(4) Yu, Bally, \& Devine (1997); Reipurth (1999);  Yu et al. (2000);} \\
\multicolumn {4} {l} { Reipurth, Rodr\'iguez \& Chini (1999); Tsujimoto et al. (2001b)}\\
\end{tabular}
\end{center}
\end{table}

A notable result is the discovery of highly absorbed X-ray
sources from the dust condensations MMS 2 (CSO 6) and MMS 3 (CSO 7) 
in the north of OMC 3 (Tsuboi et al. 2001). Since the photon statistics
is limited, the spectra are fitted with a thin thermal model by fixing the
temperature to be  3.2 keV, following the results of other sources 
in this region.  Table 1 shows the results of the two
X-ray sources.
The  absorption columns are (1-3)$\times10^{23}$ Hcm$^{-2}$, 
which are larger
than those of class Is obtained with $ASCA$ (Kamata et
al. 1997; Tsuboi et al. 2001) and $Chandra$ (Imanishi, Koyama \& Tsuboi
2001).  Consequently, these X-ray sources are really 
embedded in the cloud cores.

MMS 2 is associated with a prominent H$_2$ flow B (Yu, Bally, \& Devine 1997),
a Herbig-Haro object HH 331 (Reipurth 1999) and a 3.6-cm-radio source VLA 1
which is most likely due to a
thermal jet (Reipurth, Rodr\'iguez \& Chini 1999).  MMS 3 seems to be associated with a 
shock-excited H$_2$ flow D (Yu,
Bally, \& Devine 1997).
In addition, Yu et al. (2000) and Aso et al. (2000) found
molecular outflows of $^{12}$CO $J$ = 2-1 and  HCO$^+$  from the MMS 2--3 region.  
These facts support that MMS 2 and 3  probably contain  class 0 sources. 
In more detail, we resolve the X-ray emission at MMS 2 into three sources;
two are associated with $K$-band sources discovered with our follow-up deep IR observations 
(Tsujimoto et al. 2001).  The other X-ray source  has  no IR-counterpart, instead  an IR-jet is found. 
The hard X-ray source at  MMS 3  shows  no $K$-band counterpart even with our deep IR observations. 
Thus these two X-ray sources would be class 0s in  the cloud cores MMS 2 and 3 (Tsujimoto et al. 2001b).
\section{High Mass Young Stars}
 The giant molecular cloud Sgr B2, located at a projected distance of only 100 pc from the 
Galactic center (GC), is one of the richest star forming regions (SFR) in our Galaxy. 
 Due to its proximity to the GC, Sgr B2 is heavily obscured in the optical, even near infrared (NIR) and soft 
X-ray bands. Accordingly, the star formation activity  
has  been traced mainly with the radio and 
fur infrared (FIR) bands.  The radio continuum bands have revealed  nearly  60 ultra compact (UC) 
HII regions lying along  the north-to-south elongation, which are also traced by molecular 
lines  and FIR (Gaume et al. 1995; De Pree, Goss  \& Gaume 1998).  
Bipolar outflows are found from two of the 
HII complexes, Sgr B2 Main and North (Lis, et al. 1993).  Also clusters of 
OH, H${_2}$O and H${_2}$CO masers 
are found near the HII regions  (Mehringer, Goss 
\& Palmer 1994). Thus Sgr B2 is comprised of many clusters of high mass YSOs.
\begin{figure}[h]
\begin{center}
\hspace*{-0.5cm}
\psfig{file=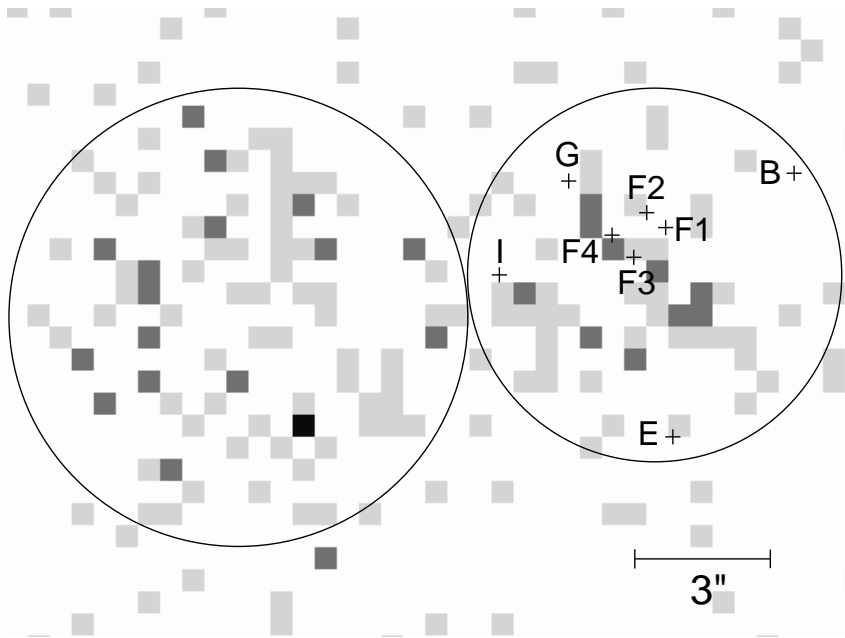,width=0.85\textwidth}
\end{center}
\caption{The  X-ray image near the HII complex Sgr B2 M. Crosses are the position
of ultra compact HII regions (Gaume et al. 1995). The X-ray spectra are
obtained from the circles (see text). 
}
\end{figure}
\begin{table}
\begin{center}
\begin{tabular}{ccccc}
\multicolumn {5} {c} {Table 2. X-ray sources in the Sgr B2 Main Complex}\\
\\ 
\hline \hline
Source$^1$ &$kT^2$  &$N_{\rm H}^3$  & $L_{\rm x}^4$ & $EW^5$ 
\\
\hline              
A	& 10 (4.1-30) 	& 4.0 (1.9-8.8)	& 8 		& 0.6	\\
B	& 4.8 ($\geq 1.1$)	& 4.0 (2.4-7.3)	& 13	&  -	\\
\hline 
\\  
\multicolumn {5} {l} {Parentheses indicate  90\% confidence range.}\\ 
\multicolumn {5} {l} {(1) A: HII regions F-I, B: The east of HII region I }\\ 
\multicolumn {5} {l} {(Gaume et al. 1995)}\\ 
\multicolumn {5} {l} {(2) Temperature in units of keV}\\
\multicolumn {5} {l} {(3) Absorption in units of 10$^{23}$ Hcm$^{-2}$}\\
\multicolumn {5} {l} {(4) Luminosity in the 2-10 keV band in units of $10^{32}$ erg s$^{-1}$}\\
\multicolumn {5} {l} {(5) Equivalent width of the 6.4 keV line in units of keV}\\
\end{tabular}
\end{center}
\end{table}

 More than a dozen X-ray sources are found from the  Sgr B2 
cloud region (Takagi, Murakami \& Koyama 2001), of which the second and third brightest (here, sources A and B)  
are located in the HII complex Sgr B2 M (Main).  
Figure 1 shows the X-ray image in the 2-10 keV band, overlaid on the positions of UC HII regions.  
Source A seems to be extended  lying  in  the UC HII regions
F and I, hence would be comprised of several point sources.  The X-ray peak
comes near the position of the brightest UC HII, F3d (De Pree, Gross and Gaume 1998)
, which is also considered to be an engine of the HC$_3$N bipolar flow
(Lis et al. 1993).  
From the HII and outflow data, this X-ray peak may correspond to a very high mass (e.g. O6, assuming ZAMS) young 
(well below $10^5$ year) star.

The X-ray components of source A  are lying along the UC HII regions  F3, F4 and G.
Figure 2 (right) shows the composite  spectrum of source A, taken from the  circle
given in figure 1 (right). 
The spectrum is nicely fitted   with  a thin thermal  model as  
is shown in table 2 and figure 2 (right).

Source B is also a complex lying in the east of the HII region I, as
is seen in the left circle of figure 1. No association with HII region may indicate
that source B is a cluster of high mass PMSSs.  
Figure 2 (left)  and table 2 show the X-ray spectrum and the
best-fit thin thermal model.  
Like source A, the best-fit temperature and 
absorption are the largest among the known stellar X-ray sources. 
This supports the hypothesis 
that sources A and B are lying at and near  the center of the Sgr B2 cloud, and demonstrate
that hard X-rays are very powerful to discover deeply embedded stars even if they 
are suffered with a large optical extinction of $A$v$\sim$200-300 $mag$.  

The absorption corrected luminosity 
in the  2-10 keV band for sources A and B are  $8\times 10^{32}$ ergs s$^{-1}$ and $13\times 10^{32}$ ergs s$^{-1}$, 
respectively.  Suppose these X-ray sources
are  complex of several high mass YSOs, then the individual luminosity would be in the order  
of $10^{32}$ ergs s$^{-1}$, 
similar to the nearest 
high mass YSO,  $\theta$ C Ori of spectral type O6-7 (Schulz et al. 2000), 
and larger than those of  younger stars, IRS 1-3 in Mon R2
(Kohno, Koyama \& Hamaguchi 2001).  The high temperature plasma  like 5-10 keV
is also found from these high mass YSOs.
Other than the HII regions F-I, no clear X-rays are found from  most of the UC HII regions;
we see only a hint of weak X-rays from the HII regions 
K (Sgr B2 North) and H (Sgr B2 South).  

A notable feature of source A is strong lines  at 6.7 keV and 6.4 keV. The former is
the K-shell transition line from  He-like irons in a thin hot plasma.  
The iron abundance is
more than 5 times of solar, which is significantly larger than that from any other high-mass SFRs
 (Kohno Koyama \& Hamaguchi 2001; Schulz et al. 2001; Yamauchi et al.1997).
The later line (6.4 keV line) may be  due to neutral or low-ionization irons 
in the dense circum/inter stellar medium irradiated by the central stars.
We note that even stronger  6.4 keV line is found from a larger region extending 
over the whole Sgr B2 cloud, which may be originated from a very luminous external X-ray source 
(Murakami, Koyama \& Maeda 2001).
The iron abundance of source B, in contrast to source A, is sub-solar.
Are the chemical abundances
really different or is the X-ray emission mechanism different with each other, 
or is there any other reason ?  In any case, how does nature make such a large
difference between these two star clusters  separated only 0.2-0.3 pc in projection ?  

\begin{figure}[t]
\begin{center}
\begin{tabular}{cc}
\psfig{file=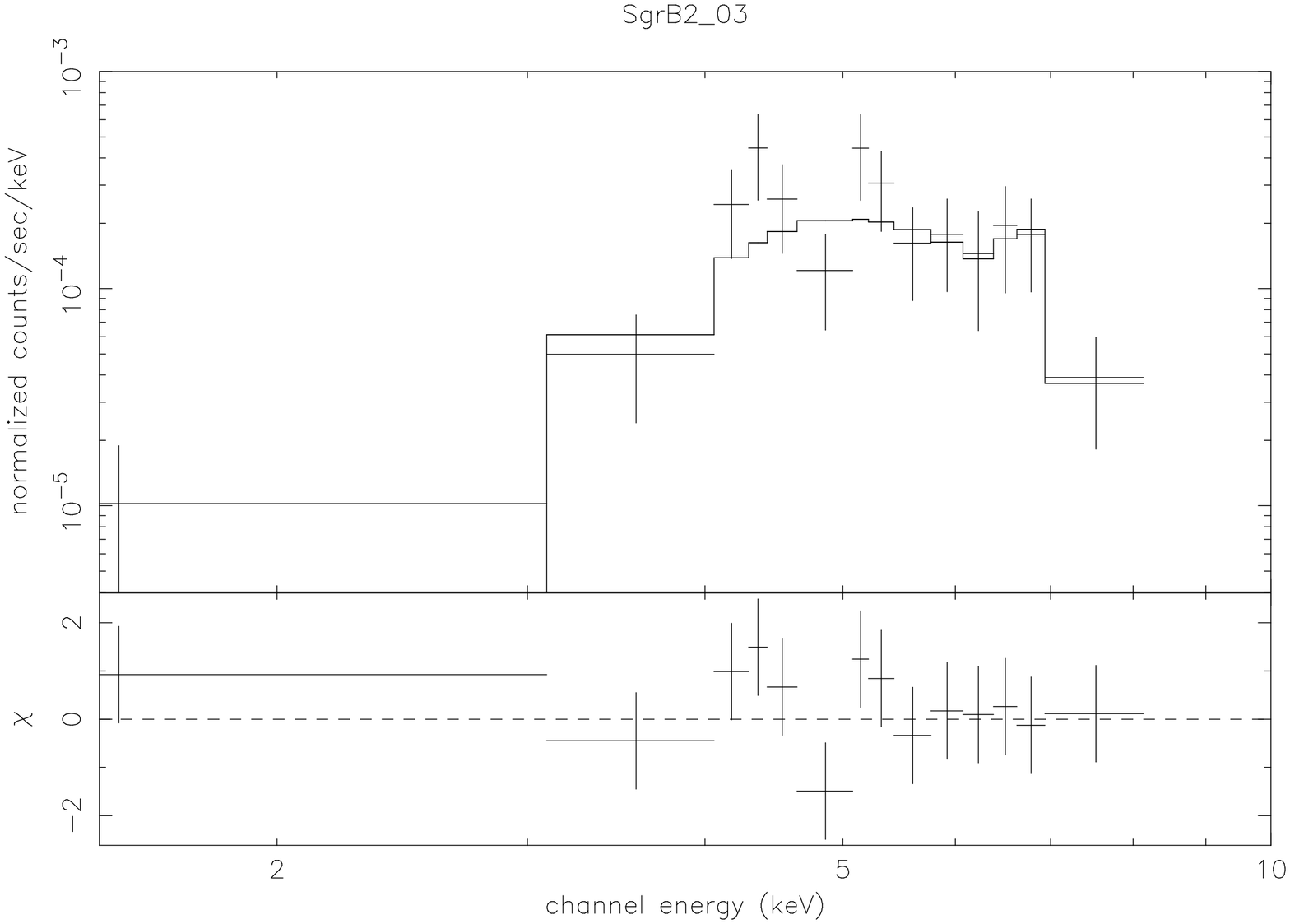,width=0.45\textwidth} &
\psfig{file=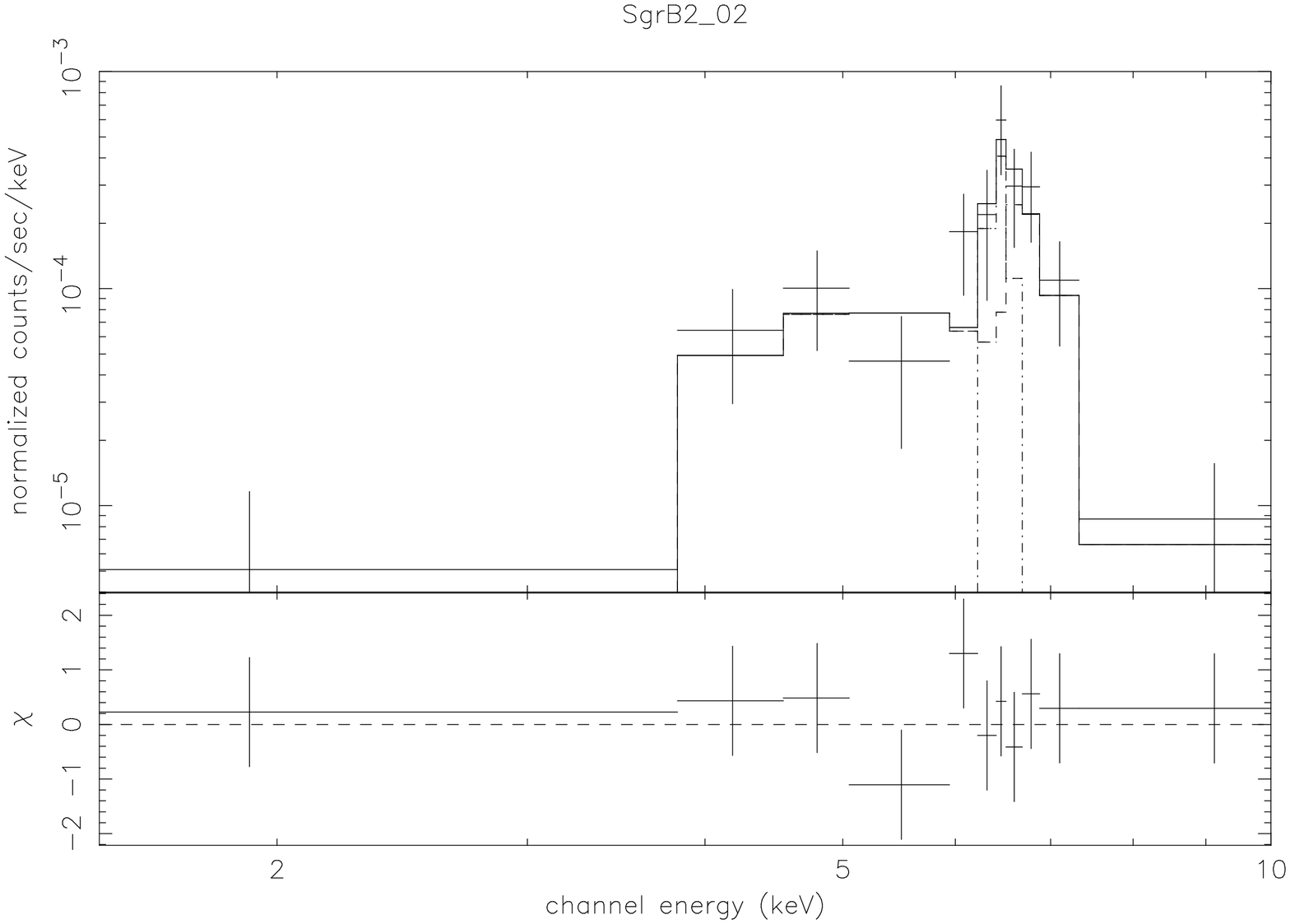,width=0.45\textwidth}
\end{tabular}
\end{center}
\caption{Right: The X-ray spectrum  and the best-fit model (solid histogram) of 
a thin thermal plasma with a 6.4 keV line (the dotted histogram) for source A. 
Left: Same as the right figure but for source B.}
\end{figure}

We have observed another high-mass SFR, the Mon R2 cloud 
located at a distance of 830 pc (Racine 1958) toward the anti-GC direction.
Figure 3 shows the ACIS-I image of  the central
region of the  cloud.  We see heavily absorbed X-rays  
from high mass YSOs,  IRS 1SW, IRS 2, IRS 3NE and ${\rm a_S}$
\footnote {For the IR sources, we refer the original naming by 
Beckwith et al. (1976)}. The X-ray features with the best-fit thin thermal 
model  are listed in table 3.

IRS 1 is located in a compact HII region in an IR shell 
(Massi, Felli, \& Simon 1985). 
It has been resolved into two IR stars, IRS 1SW and NE (Howard, Pipher \& Forrest 1994). IRS 1SW is exciting 
the compact HII region with
a predicted luminosity of
$\geq 10^4L_\odot$,   hence would be  a B0 of zero-aged main sequence (ZAMS).
The IR source ${\rm a_S}$ has been less studied, hence is not well known. 
It is associated with a small IR nebulosity and has
a similar IR spectrum and bolometric luminosity to those of IRS 1SW, 
hence may be a similar mass star, early B or late O type.  
Since ${\rm a_S}$ has no HII region, it may be younger than IRS 1SW. 
We find that IRS1 SW and ${\rm a_S}$ are strong and heavily absorbed X-ray sources, 
consistent with their $H-K$ values. 
These show  a high temperature plasma of $\sim$2-3 keV
and a rapid time variability including  flare-like events,
in contrast to high mass MSSs with lower temperature plasma 
of $\leq$1~keV and  relatively  stable light curve.
We thus suspect that IRS 1SW and ${\rm a_S}$ have higher magnetic  activity
than that of stellar wind.
The  X-ray luminosity is $\sim 10^{31}$erg s$^{-1}$, which
is significantly higher   than that of low mass YSOs.

IRS 2 is an illuminating source of the IR shell. The bolometric luminosity is
$\sim 0.5 \times 10^4L_\odot$, hence would be  a high mass YSO of $8-10M_\odot$ 
(Howard, Pipher \& Forrest 1994).
\begin{figure}[t]
\begin{center}
\hspace*{-0.5cm}
\psfig{file=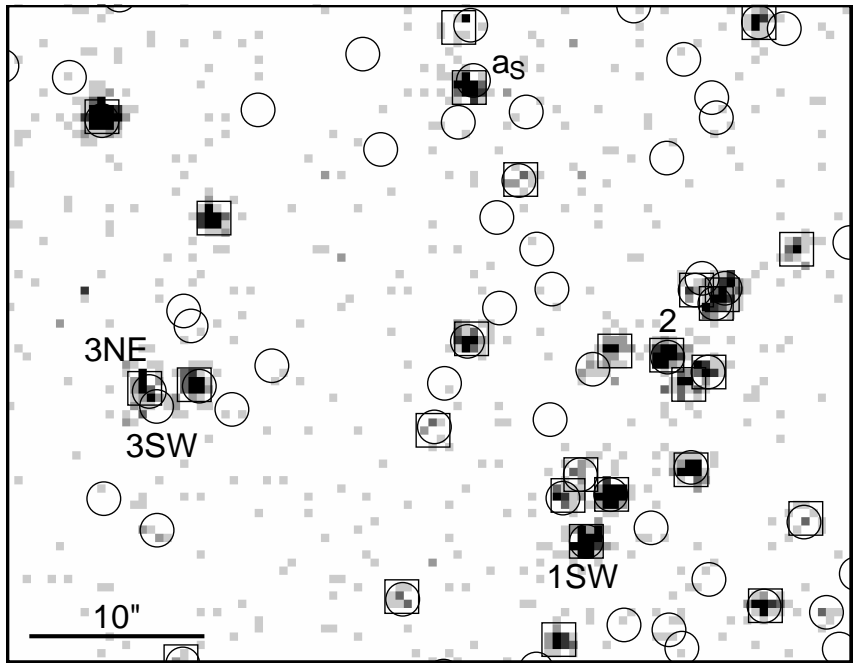,width=0.85\textwidth}
\end{center}
\caption{Expanded view of the Mon R2 cloud core of a $50'' \times 50''$ region. 
Circles and squares indicate the positions of infrared and X-ray sources, respectively
(Carpenter et al. 1997; Kohno, Koyama \& Hamaguchi 2001).
High mass  YSOs  are shown  with the labels of  ${\rm a_S}$, (IRS) 1SW, (IRS) 2, (IRS) 3NE and 3SW.}
\end{figure}
The infrared spectrum of IRS 2 (also IRS 3)
shows deep absorption features in the water-ice and silicate bands,
which is a signature  of a cluster of several young embedded sources 
(Smith, Sellgren \& Tokunaga  1989;  Sellgren, Smith \& Brooke 1994; Sellgren et al. 1995).
Thus IRS 2 may contain high mass star(s) in younger phase  than IRS 1 at ZAMS.  
Our $Chandra$ observation  reveals that
IRS 2 has the highest plasma temperature ($\sim 10$ keV) and the largest absorption 
column density ($\sim 10^{23}$ Hcm$^{-1}$) among the bright sources in the Mon R2 cloud. 
Large absorptions have been  reported from embedded low mass stars, like class I or 0 
protostars (Koyama et al. 1996; Imanishi, Koyama \& Tsuboi 2001).
Therefore, the heavily absorbed hard X-ray source at IRS 2 is very likely to be
a high mass YSO.
Since the X-ray time variation is long enough to be rotational 
modulation,  it may be arisen from the interaction  of stellar wind with
magnetosphere (e.g. Gagne et al, 1997).
However,  IRS 2  has neither HII region, hence no strong UV field, 
nor strong stellar wind.  Furthermore  {\it ASCA}  detected  a big flare from the position of IRS 2, 
although possible source confusion can 
not be excluded (Hamaguchi, Tsuboi \& Koyama 2000). 
Therefore, X-rays from IRS 2 are  likely  magnetic origin. 

IRS 3, the brightest source in the near- and mid-IR bands in Mon R2, is another active star forming 
site. The presence of ${\rm H}_{2}$O and OH masers and
a well-developed molecular outflow indicate that IRS 3 is till in a  phase
of dynamical mass accretion.
\begin{table}[h]
\begin{center}
\begin{tabular}{cccc}
\multicolumn {4} {c} {Table 3. High mass YSOs in Mon R2}\\ 
\\ 
\hline \hline
Name		&$kT^1$  	& $N_{\rm H}^2$   	& $L_{\rm x}^3$ \\
\hline              
IRS 1SW 	& 1.9(1.3-2.8)	& 4.7(3.5-6.4)		& 1 (0.6-2.0)		\\
${\rm a_S}$	&2.6(1.6-6.2)	&6.0(4.0-9.2)		&0.6(0.4-1.6)		\\
IRS 2		&10.9($\geq 2.0$)&9.1(5.7-17)		&0.6(0.5-2.0)	  	\\
\hline
\\
\multicolumn {4} {l} {Parentheses indicate 90\% confidence range.}\\
\multicolumn {4} {l} {(1) Thin thermal temperature in units of keV}\\
\multicolumn {4} {l} {(2) Absorption in units of 10$^{22}$ Hcm$^{-2}$}\\
\multicolumn {4} {l} {(3) The 0.5-10 keV band luminosity in units of $10^{31}$~ergs~s$^{-1}$}\\
\end{tabular}
\end{center}
\end{table}
From the velocity and size of the outflow, the age is estimated to be
$\sim10^5$ years, hence would contain a youngest high-mass  star in the cloud. 
IRS 3 has been resolved into two sources IRS 3NE and 3SW, of which the former
(IRS 3NE) exhibits heavily absorbed hard X-rays, hence is  a 
candidate of a high mass YSO embedded in the cloud core.
The infrared polarimetry reveals  that the
interstellar magnetic field is compressed from  the neighbor of the GMCs
(e.g., Yao et al. 1997).  If the magnetic field  is connecting  the stellar surface and the accreting gas disk, 
it can be twisted, amplified and reconnected
by the differential rotation. Then a high temperature plasma is produced 
by the release of the magnetic energy (Tsuboi et al. 2000;  Montmerle, et al. 2000).  The outflow found from  
IRS 3  is  also  explained  by this scenario (Hayashi, Shibata \& Matsumoto 1996). 

Since the majority of stars in the universe
seem to form in binary pairs, one may argue that these high mass  stars are  
binaries with a low mass companion, and
the low mass YSO  may be responsible for the X-ray emission.
 The X-ray luminosity of these high mass YSOs, $\sim
10^{30-31}$~ergs~s$^{-1}$, is however 
significantly larger than that of typical low mass YSOs of $\sim
10^{28-29}$~ergs~s$^{-1}$.  Thus contribution of low mass companion, if any,
may be a small fraction of the bulk X-rays observed from the high mass YSOs. 

 Here we propose a working hypothesis  for the study of X-ray evolution of
high mass YSOs; high mass  stars produce variable and hard (2-10 keV) X-rays due to the magnetic
activity in PMS (source B in Sgr B2, IRS 2 and IRS 3NE in Mon R2). 
It continues until ZAMS (source A in Sgr B2, IRS 1SW in Mon R2), 
then gradually decline to  stellar wind dominant activity. 
Finally, in the main sequence phase, high mass stars 
predominantly emit soft ($\leq$1 keV 
temperature) X-rays originated from  the strong stellar wind (e.g. the Orion Trapezium stars).
\section{Young Brown Dwarfs}
$Chandra$ made two deep exposure observations on the center region of the  $\rho$ Oph cloud
core A and cores B-F.   We search for  X-ray emission from the brown dwarf catalogs
selected with the IR observations on these 
regions (Cushing, Tokunaga \& Kobayashi 2000; Wilking, Greene \& Meyer 1999).  
Two  out of eight bona fide brown dwarfs (BDs) and 
five out of  ten candidate brown dwarfs (CBDs) are found to be X-ray sources 
(Imanishi, Tsujimoto \& Koyama 2001) \footnote 
{We call a bona fide brown dwarf (BD)
for stars with the stellar mass well below the hydrogen burning limit of 0.08$M_\odot$
and  a candidate brown dwarf (CBD) for stars in a transition mass region}.
The X-ray detection rates are, therefore 25\% and 50\% for BDs and CBDs, respectively.  
\begin{table}[h]
\begin{center}
\begin{tabular}{cccccc}
\multicolumn {6} {c} {Table 4. X-rays from Brown Dwarfs} \\ 
\\
\hline \hline
Name$^1$ &Sp.type$^2$ & $kT$$^3$  & $N_{\rm H}$$^4$ &$L_{\rm x}$$^5$&log($L_{\rm x}$/$L_{\rm bol}$)$^6$\\
\hline              
\multicolumn {6} {l} {Bona fide Brown Dwarf --------------------------- } \\
GY141 	&M8	& -	 	& -		&0.32	&-3.5 	\\
GY310 	&M8.5	& 1.7(0.9-2.2)	& 0.8(0.5-1.5)	&12	&-3.2	\\
\multicolumn {6} {l} {Candidate Brown Dwarf --------------------------- } \\
 GY31	&M5.5	& 2.2(1.7-2.9)	& 5.9(5.1-7.2)	&120	& -3.4	\\
 GY37	&M6	& -		& -		&0.76	& -4.1	\\
 GY59	&M6	&2.5($\geq 1.0$)& 1.4(0.5-3.0)	&3.4	& -3.8	\\
 GY84	&M6	& -		& -		&1.1	& -4.5	\\
 GY326	&M4	& 0.9(0.7-1.2)	& 2.3(1.6-3.0)	&36	& -3.3	\\
\hline 
\\
\multicolumn {6} {l} {Temperatures are fixed to 2.0 keV, for GY141, GY37 and GY84.}\\ 
\multicolumn {6} {l} {Parentheses indicate 90\% confidence range.}\\
\multicolumn {6} {l} {(1), (2), (6) References: Cushing, Tokunaga \& Kobayashi (2000);}\\
\multicolumn {6} {l} {Wilking, Greene \& Meyer (1999)}\\
\multicolumn {6} {l} {(2) Spectral type, uncertainty is $\pm1.5$.}\\
\multicolumn {6} {l} {(3) Thin thermal temperature in units of keV}\\
\multicolumn {6} {l} {(4) Absorption in units of 10$^{22}$ Hcm$^{-2}$}\\
\multicolumn {6} {l} {(5) The 0.5-9 keV band luminosity in units of $10^{28}$~ergs~s$^{-1}$}\\
\multicolumn {6} {l} {(6) Luminosity ratio between X-ray and bolometric}\\
\end{tabular}
\end{center}
\end{table}
Table 4 shows the list of X-ray detected brown dwarfs together with the 
best-fit parameters of a thin thermal model.
The luminosity ratio of X-ray
($L_{\rm x}$) to bolometric ($L_{\rm bol}$), $L_{\rm x} /L_{\rm bol}$ 
, lies 
between $10^{-3} - 10^{-5}$, which is similar to those of low-mass pre-main-sequence stars 
(e.g., Imanishi, Koyama \& Tsuboi 2001) and dMe stars (Giampapa et al. 1996).
For the X-ray non-detected BDs and CBDs, the upper limits of $L_{\rm x} /L_{\rm bol}$
are also scattered around $10^{-3} - 10^{-5}$, comparable with or slightly lower 
than those of X-ray detected samples. This leads us to suspect that the
X-ray non-detected  BDs and CBDs may also emit X-rays  near  at the current 
sensitivity limit of $Chandra$.
 
One BD (GY310) and three CBDs (GY31, GY59 and GY326) are bright enough, 
hence reliable X-ray spectra and light curves can be made for the first time. 
The spectra are fitted with a thin thermal plasma model of $\sim$2 keV temperature. 
Solar-like flares are  detected from 2 CBDs (GY31 and GY59). These X-ray features
are similar to those of low mass stars. 
 Together with the high $L_{\rm x} /L_{\rm bol}$ value, we suggest that X-rays 
from these sub-stellar objects (BDs and CBDs) are 
attributable to magnetic activity similar to low-mass stellar objects. 

A debatable issue is the mechanism of magnetic field amplification. Brown dwarfs are 
fully convective, hence the standard dynamo mechanism may not work. This situation
is the same as low mass protostars.  One possible scenario is the magnetic arcade 
scenario connecting  a star and disk, which can also be applied for the X-rays from high mass 
YSOs (see section 3), although stellar structures may be largely different (Tsuboi et al. 2000;  
Montmerle et al. 2000). 
If BDs and CBDs emit  X-rays with this  mechanism, X-ray activity should vanish when 
the disk disappears
as  BDs and CBDs evolve, but our observations show no clear evidence for the X-ray evolution with ages.
We therefore conclude that an accretion  disk is not a key factor for the X-ray emissions from 
BDs and CBDs. 
Another type, such as turbulent-driven  dynamo 
(Durney, De Young, \& Roxburgh 1993), should be involved.

\acknowledgements
The author expresses his thanks to Drs. Y. Tsuboi, K. Hamaguchi,  H. Murakami
K. Imanishi, M. Tsujimoto, M. Kohno and S. Takagi for their collaborations.

\end{document}